\documentclass[aps,prl,twocolumn,nofootinbib,groupedaddress,superscriptaddress
,longbibliography]{revtex4-2}

\usepackage{amsmath,amssymb}
\usepackage{multirow}
\usepackage{bm}
\usepackage{mathtools}
\usepackage{dsfont}
\usepackage{amsfonts}
\usepackage[normalem]{ulem}
\usepackage{url}
\usepackage{wasysym}
\usepackage{bbm}
\usepackage{subfigure}
\usepackage[section]{placeins}

\usepackage[colorlinks=true,linkcolor=magenta,citecolor=magenta,hypertexnames=false]{hyperref}

\def\ba#1\ea{\begin{align}#1\end{align}}
\def\bg#1\eg{\begin{gather}#1\end{gather}}
\def\bpm{\begin{pmatrix}}
\def\epm{\end{pmatrix}}

\def\Fig#1{Fig.~\ref{#1}}

\newcommand{\magenta}[1]{\textcolor{magenta}{#1}}

\newcommand{\bea}{\begin{eqnarray}}
\newcommand{\eea}{\end{eqnarray}}

\allowdisplaybreaks

\begin{document}
\title{Superradiant strongly correlated quantum states in cavity Hubbard model}

\author{Kang Wang}
\affiliation{Beijing National Laboratory for Condensed Matter Physics \& Institute of Physics, Chinese Academy of Sciences, Beijing 100190, China}
\affiliation{Department of Physics and Astronomy, University of Pittsburgh, PA 15213, USA}
\author{Wei-Xuan Chang}
\affiliation{Beijing National Laboratory for Condensed Matter Physics \& Institute of Physics, Chinese Academy of Sciences, Beijing 100190, China}
\affiliation{Institute for Advanced Study, Tsinghua University, Beijing 100084, China}
\author{Cheng-Yu Bi}
\affiliation{Beijing National Laboratory for Condensed Matter Physics \& Institute of Physics, Chinese Academy of Sciences, Beijing 100190, China}
\affiliation{College of Science, Northeast Forestry University, Harbin 150040, China}
\author{Zi Cai}
\email{zcai@sjtu.edu.cn}
\affiliation{Wilczek Quantum Center and Shanghai Research Center for Quantum Sciences, School of Physics and Astronomy, Shanghai Jiao Tong University, Shanghai 200240, China}
\affiliation{Key Laboratory of Artificial Structures and Quantum Control, School of Physics and Astronomy, Shanghai Jiao Tong University, Shanghai 200240, China}
\author{Zi-Xiang Li}
\email{zixiangli@iphy.ac.cn}
\affiliation{Beijing National Laboratory for Condensed Matter Physics \& Institute of Physics, Chinese Academy of Sciences, Beijing 100190, China}
\affiliation{University of Chinese Academy of Sciences, Beijing 100049, China}


\begin{abstract}
In cavity quantum materials, entangling strongly correlated electrons with quantum light provides a unique opportunity to explore novel quantum phases and phase transitions absent in conventional solid-state materials. In this study, we develop a sign-problem-free fermion-photon hybrid  Quantum Monte Carlo (QMC) algorithm, and use it to systematically investigate the ground-state phase diagram of a two-dimensional cavity Hubbard model. It is shown that the interplay between the electron correlation and photon condensation gives rise to intriguing quantum phases ({\it e.g.}  superradiant antiferromagnetic and chiral/$\pi$-flux states), and different quantum phase transitions, such as a first-order superradiant phase transition and a continuous phase transition with  Gross-Neveu universality class. The methodology can be readily generalized to more complicated cavity strongly correlated models. 
\end{abstract}

\maketitle

\let\oldaddcontentsline\addcontentsline
\renewcommand{\addcontentsline}[3]{}
The integration of quantum matter with cavity quantum electrodynamics (QED) has given rise to an emerging field at the intersection of condensed matter physics and quantum optics, now termed ``cavity quantum materials"~\cite{Sentef2022}, which enables access to the ultrastrong and deep-strong light-matter coupling regimes where the properties of quantum materials can be drastically modified~\cite{Bloch2022Nature,Ebbesen2021ScienceReview,Solano2019RMP,Ruggenthaler2018NRCReview}. Typical examples in this regard include the cavity-enhanced/altered superconductor~\cite{Sentef2018,Schlawin2019,Galitski2019PRLCavitySC,Chakraborty2021,Rubio2024PNAS,Basov2026NatureCavityAlteredSC} and topological phases mediated by vacuum fluctuations~\cite{Faist2022ScienceCavity,Enkner2025,Chiocchetta2021,Wei2025}. The situation becomes particularly intriguing when the cavity quantum material itself is a strongly correlated system, where the interplay between strong electronic correlations and vacuum fluctuations of the cavity mode can give rise to phenomena beyond those anticipated from either effect alone.

The Hubbard model, a cornerstone of strongly correlated electron physics, captures the competition between kinetic energy of electrons and local Coulomb repulsion to explain emergent phenomena ranging from Mott insulator to high-temperature superconductivity within a minimal framework~\cite{Arovas2022,Qin2022}. In a cavity system, this naturally leads to the consideration of its extension—the Hubbard model coupled to a single photon mode—as the fundamental starting point for investigating strongly correlated physics within a cavity environment. However, a theoretical understanding of such a cavity Hubbard model poses a formidable many-body numerical challenge, which calls for an equal footing treatment of the non-perturbative light-matter coupling and strong Coulomb repulsion.  Despite pioneering theoretical efforts~\cite{Li2020, Demler2020PRX,23DMRG,Tokatly2021,Rao2023,Georges2019PRL}, to date, an unbiased non-perturbative study of a strongly correlated cavity Hubbard model in dimensions higher than one has remained fundamentally lacking.

\begin{figure}[!htbp]
    \centering
    \makebox[0.8\linewidth][l]{%
        \includegraphics[width=0.8\linewidth]{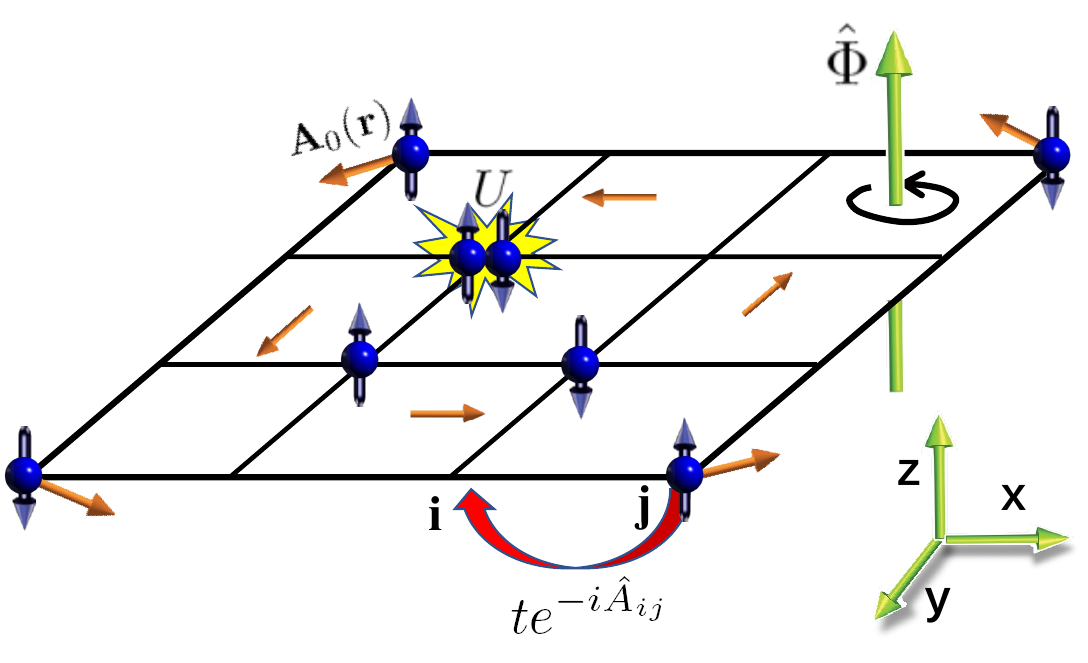}%
        \put(-200,110){\large\textbf{(a)}}%
    }
    \vspace{-5pt}
    \makebox[0.8\linewidth][l]{\includegraphics[width=0.8\linewidth]{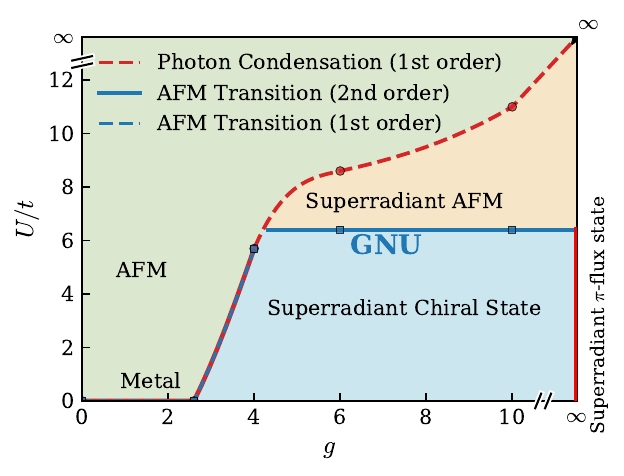}%
        \put(-200,140){\large\textbf{(b)}}%
    }
    \caption{(a) Schematic illustration of a two-dimensional Hubbard model embedded in a gyrotropic cavity. (b)Ground-state phase diagram of the cavity Hubbard model in terms of light-matter coupling $g$ and Hubbard interaction $U/t$. Dashed and solid denote first-order and second-order AFM transitions, respectively. The label GNU marks the second-order phase transition with  Gross-Neveu universality class.}
    \label{fig:diagram}
\end{figure}

To address this issue, in this study, we develop a Quantum Monte Carlo (QMC) algorithm for photon-fermion hybrid systems, and apply it to investigate a half-filled two-dimensional Hubbard model embedded in a gyrotropic cavity, as shown in Fig.\ref{fig:diagram} (a).  This method treats fermions and photons on an equal footing, and is free of the negative sign problem, thereby enabling an unbiased and well-controlled investigation of the ground state phase diagram of this cavity Hubbard model. It is shown that the interplay between the electron correlation and the photon condensation leads to rich quantum phases and phase transitions (see Fig.\ref{fig:diagram} (b)).  In the absence of repulsive interaction,  the system undergoes a first-order superradiant phase transition, manifesting as photon condensation that induces a spatially uniform magnetic field and gives rise to an emergent chiral or $\pi$-flux fermionic state. With Coulomb repulsion, a superradiant antiferromagnetic (AFM) phase emerges in a regime bounded by a first-order superradiant transition and a continuous transition with the Gross-Neveu universality class. Such a superradiant AFM phase is characterized by the coexistence of an emergent magnetic flux and antiferromagnetic order- a phenomenon rarely observed in conventional solid-state materials. 


\magenta{\it Model and method|}
We consider a two-dimensional model of interacting electrons coupled to a single-mode photon field inside a gyrotropic cavity defined in a $L\times L$ square lattice. The Hamiltonian is given by:
\begin{equation}
\hat{H} = \hat{H}_e+ \hat{H}_{ph}\label{eq:Ham}
\end{equation}
where $\hat{H}_e$ and  $\hat{H}_{ph}$  depict the Hamiltonian of fermions and the cavity mode respectively, which reads: 
\bea
\nonumber \hat{H}_e &=& -t\sum_{\left<ij\right>,\sigma} (e^{-i\hat{A}_{ij}}\hat{c}_{i\sigma}^\dagger \hat{c}_{j\sigma} + h.c.)
\\&+& U\sum_i \hat{n}_{i\uparrow} \hat{n}_{i\downarrow} -\mu \sum_{i\sigma}\hat{n}_{i\sigma}  ,\nonumber\\
 \hat{H}_{ph} &=&\frac{\hat{P}^2}{2m}+\frac{1}{2}m\omega_c ^2 \hat{X}^2.
\label{model}
\eea
where the summation $\left<ij\right>$ is over the adjacent sites on the square lattice, $\hat{c}_{i\sigma}^\dagger(\hat{c}_{i\sigma})$ creates (annihilates)  a fermion with spin $\sigma$ on site $i$, and $\hat{n}_{i\sigma}= \hat{c}_{i\sigma}^\dagger \hat{c}_{i\sigma}$.   $t$ is the single-particle hopping amplitude, and  $U>0$ denotes the strength of the on-site repulsive interactions between the fermions. We focus the study on the square lattice at half filling, thus the chemical potential of the fermion $\mu$  is set to be $\mu=U/2$ throughout this paper to guarantee the half-filled condition. The cavity photon mode is treated as a harmonic oscillator with position operator $\hat{X}$ and momentum operator $\hat{P}$, with a cavity frequency $\omega_c$, which is  fixed as $\omega_c=t$ throughout this work. Both the lattice constant and the charge of the fermion are set to be unit for simplicity.  We also set $t$ as the unit for all quantities with dimensions of energy.

The light-matter interaction is incorporated via the Peierls substitution~\cite{Luttinger1951}, where the hopping amplitude acquires a phase dependent on the vector potential: $t_{ij}\to t_{ij}e^{-i\int_{r_i}^{r_j}d\mathbf{r}\cdot \hat{\mathbf{A}}(\mathbf{r}) }$. $\hat{\mathbf{A}}(\mathbf{r})$ is the vector potential operator of the photon mode in gyrotropic cavity~\cite{Rubio2021NMReview,Fausti2023Nature,Rubio2023npjQM}, which is defined as:
\begin{equation}
\hat{\mathbf{A}}(\mathbf{r}) = \sqrt{\frac 2 N}g~\hat{X}~ \mathbf{A}_0(\mathbf{r}),
\end{equation}
The spatially uniform operator $\hat{X}$ arises from the quantization of the single cavity mode.  $g$ is a dimensionless parameter characterizing the light-matter coupling strength.  The vector $\mathbf{A}_0(\mathbf{r})$ retains the spatial dependence of the vector potential, and  we adopt the symmetric gauge $\mathbf{A}_0(\mathbf{r})=(-y/2, x/2)$ with $(x, y)$ being the spatial coordinates in the x-y plane. $\hat{\mathbf{A}}(\mathbf{r})$ has a nonzero curl, thus such a cavity mode generates a fluctuating magnetic flux in each plaquette, which is described by the operator:
\begin{equation}
\hat{\Phi}=\int_\square  dx dy \nabla \times \hat{\mathbf{A}}=\frac{2g}{\sqrt{N}}\hat{X}
\end{equation}
where $N=L^2$ and the integral is over a plaquette in the square lattice. The prefactor $\sqrt{1/N}$  guarantees that the flux in each plaquette doesn't depend on the system size once the photons condense ($\langle\hat{X}\rangle\sim \sqrt{N}$). A similar non-interacting fermionic model defined in a two-leg ladder system  has been investigated by a Density-Matrix-Renormalization-Group method~\cite{23DMRG}.

To investigate such a hybrid model, we perform a determinant quantum Monte Carlo simulation with local updates for the discrete Hubbard-Stratonovich auxiliary fields~\cite{BSS,AssaadReview,lzx2019review}, complemented by the  hybrid Monte Carlo updates~\cite{HQMC87,SmoQyDQMC.jl,18Assaad,Cohen-Stead2022PRE} for the photon coordinate $X(\tau)$ with $\tau$ being the imaginary time. This algorithm  allows us to simultaneously update the spatial-time auxiliary fields configurations of the fermions and the coordinate configuration $X(\tau)$ of the photons. By formulating the algorithm in the coordinate basis $X(\tau)$ of photon instead of its  occupation number basis, we avoid truncating the photon Hilbert space. This is particularly important in the presence of photon condensate, where the occupation number of  photons are macroscopically large.  For a half-filled square lattice, our model is free from the notorious sign problem, enabling unbiased quantum Monte Carlo simulations for large system with system size up to $L=16$. We choose open boundary condition in both directions. More details about the QMC algorithm and the sign-problem-free condition have been provided in the Supplementary Materials (SM)~\cite{re:supp}.

\begin{figure}[!htbp]
    \centering
    \includegraphics[width=\linewidth]{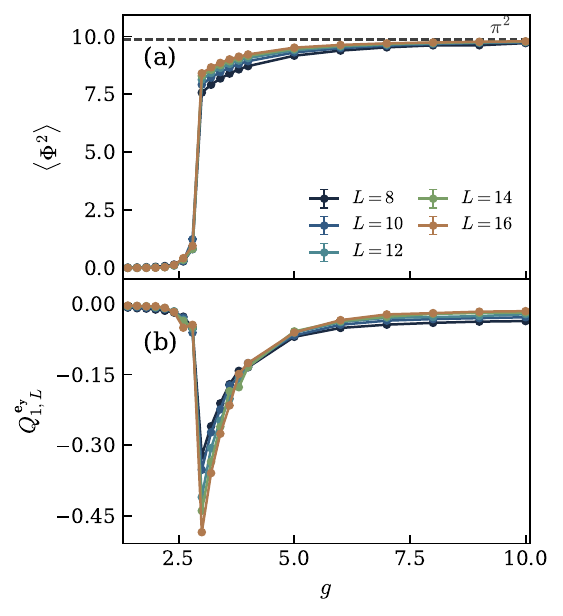}
    \caption{
     Superradiant phase transition at $U=0$.
    (a) Squared effective gauge flux $\langle \hat{\Phi}^2 \rangle=\frac{g^2}{N}\langle \hat{X}^2 \rangle$ per plaquette as a function of  light-matter coupling $g$. The discontinuous jump at $g_c\approx 2.8$ signals a first-order photon condensation transition.
    (b) Edge current correlation function. The finite negative value for $g>g_c$ indicates chiral edge currents induced by the emergent magnetic flux. }
    \label{Dirac}
\end{figure}

\magenta{\it Symmetry analysis and superradiant phase transition|} Before presenting our numerical results, we first analyze the symmetry of the Hamiltonian(\ref{eq:Ham}).  It is easy to check that at half-filling ($\mu=U/2$), the Hamiltonian in Eq.~(\ref{eq:Ham}) is invariant under the combination of a particle-hole (PH) transformation on the fermions: $\hat{c}^\dagger_{i\sigma}\to (-1)^i\hat{d}_{i\sigma}$, and a parity transformation on the photon field: $\hat{X}\to -\hat{X}$. In a light-matter coupling system,  the condensate of photon ($\langle\hat{X}\rangle\neq 0$) which spontaneously breaks the combined symmetry~\cite{Dicke1954}, leads to a superradiant phase transition which has been realized in ultracold atom experiments~\cite{Esslinger2010Nature,Esslinger2013RMP}, and has drawn considerable attention in electronic systems recently~\cite{MacDonald2012PRL,Basko2019PRL,Mora2020PRL,Andolina2020PRB}. A no-go theorem posits that photon condensation is prohibited in presence of spatially uniform vector potential~\cite{SuperradianceA2term,Superradiancenogo,CommentPRL,Ciuti2010NC,MacDonald2019PRB}, while the coupling in a gyrotropic cavity circumvents these constraints, thereby permitting the emergence of a genuine photon condensate.

\magenta{\it Superradiant chiral and $\pi$-flux phases|} We start with the non-interacting case ($U=0$). To identify the superradiant phase transition  in a finite system, we calculate the average value of $\hat{\Phi}^2$ (since $\langle\hat{\Phi}\rangle=0$ for a finite system).   The dependence of $\langle\hat{\Phi}^2\rangle$ as a function of coupling strength $g$ is plotted in \Fig{Dirac}(a), from which we can see that $\langle\hat{\Phi}^2\rangle$  undergoes a sudden jump from zero to a finite value at a critical value of $g_c\approx 2.8$, indicating a first-order quantum phase transition. This phase transition can be understood as a competition between the energies of fermions and photons: although an emergent magnetic flux will increase the energy of photons, it indeed lowers the kinetic energy of fermions.  This becomes evident in the strong-coupling limit ($g \to \infty$), where the flux per plaquette tends to $\pi$ (see \Fig{Dirac}(a)). Consequently, the system maps onto  a $\pi$-flux phase on the square lattice, whose low-energy excitations are described by Dirac fermions.  For non-interacting electrons on a half-filled square lattice, the ground-state energy is minimized in the presence of a uniform $\pi$-flux per plaquette, as dictated by Lieb's theorem~\cite{LiebFlux94}.

In the thermodynamical limit, such a superradiant chiral state with a magnetic flux per plaquette other than $0$ or $\pi$ breaks the time reversal symmetry, thus hosting a chiral charge current on the edge. To verify this point numerically in a finite system, we calculate the current-current correlation between the bonds on the opposite vertical edges, which is defined as:
\begin{equation}
Q^{\mathrm{y}}_{\mathrm{edge}} = \frac{1}{L_y^2}\sum_{y_1, y_2}\langle \hat{J}_{(1,y_1)}^{\mathbf{e_y}} \hat{J}_{(L_x,y_2)}^{\mathbf{e_y}} \rangle,
\end{equation}
where $\hat{J}_{(x,y)}^{\mathbf{e}_y}$ is the current operator defined on the vertical bond $[(x,y),(x,y+1)]$. As shown in \Fig{Dirac}(b), the quantity $Q^{\mathrm{y}}_{\mathrm{edge}}$ jumps from zero to a finite value at $g =  g_c$, with its negative sign signaling opposing currents on the two edges. 
Upon further increasing $g$,  the current correlation decays and vanishes as $g \to \infty$, where time-reversal symmetry is restored in the $\pi$-flux phase thus  the edge current disappears.


\magenta{\it Superradiant anti-ferromagnetic phase|}  Next, we explore the effect of electronic interactions (Hubbard repulsion $U$) on photon condensation and the cavity-induced chiral and $\pi$-flux phases. In a half-filled square lattice without cavity, the Hubbard interaction promotes AFM order. Our interest lies in the interplay between this interaction-driven AFM order and the cavity-induced photon condensation, and the competition between them is expected to yield a rich ground-state phase diagram.

The onset of AFM order can be identified by a peak in the spin structure factor \( S(\mathbf{q}) \):
\begin{equation}
S(\mathbf{q})=\frac{1}{N^2}\sum_{ij} e^{i\mathbf{q}\cdot(\mathbf{i}-\mathbf{j}) }\langle (\hat{n}_{i\uparrow}-\hat{n}_{i\downarrow})(\hat{n}_{j\uparrow}-\hat{n}_{j\downarrow})\rangle,
\end{equation}
at the wavevector \( \mathbf{q} = (\pi, \pi) \). For a precise determination of the critical point, we analyze the scaling of the dimensionless correlation-length ratio:
\begin{equation}
R_s = \frac{S(\pi, \pi)}{S\left(\pi + \frac{2\pi}{L}, \pi + \frac{2\pi}{L}\right)}.
\end{equation}
 In the long-range ordered phase, \( R_s \) increases with system size \( L \), while in the short-range disordered phase, it decreases. At a continuous phase transition, the system is scale-invariant, causing curves of \( R_s \) for different \( L \) to intersect at a single critical point. This crossing provides a robust and highly accurate signature of the quantum phase transition.
\begin{figure}[!htbp]
    \centering
    \includegraphics[width=\linewidth]{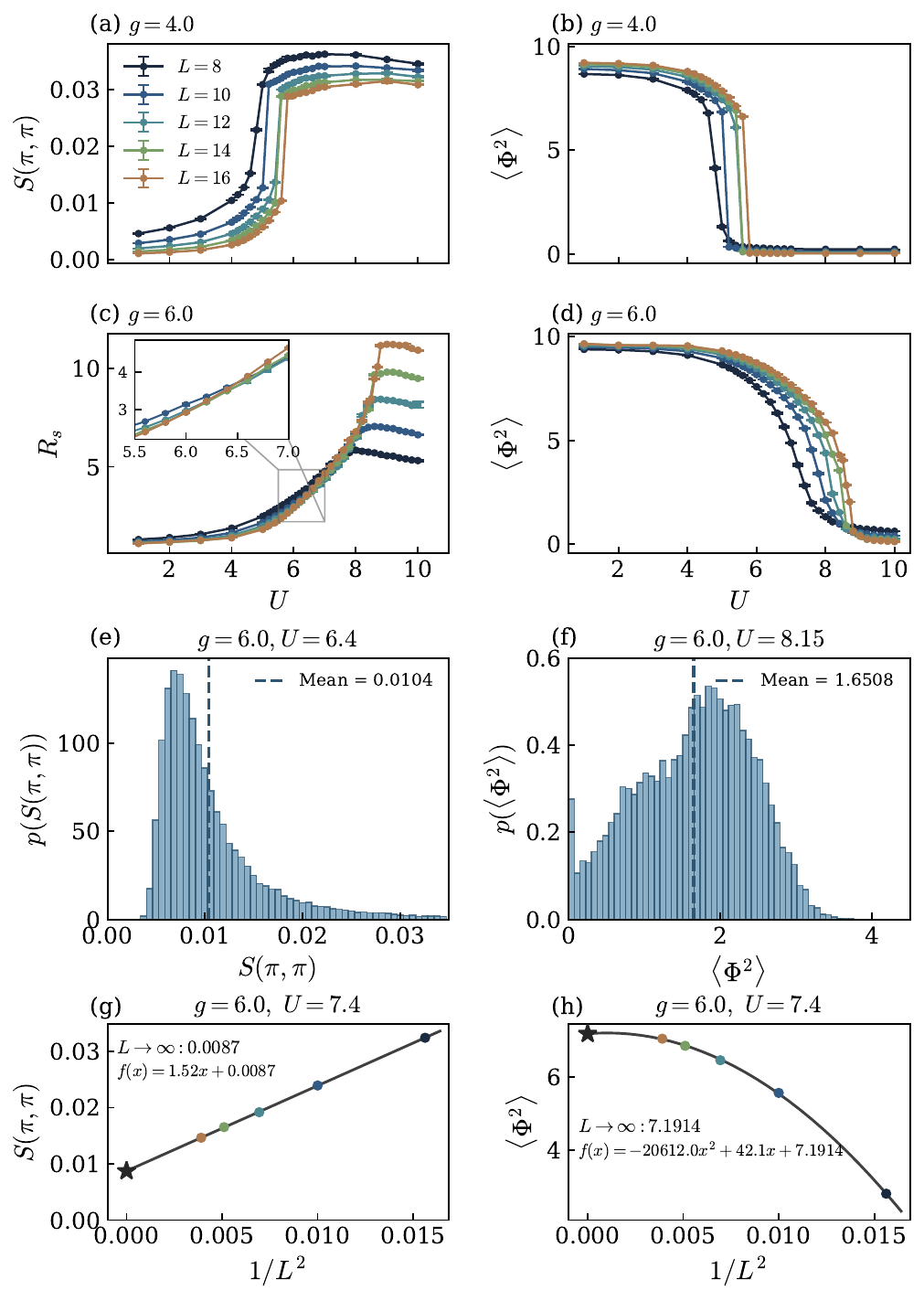}
    \caption{
    Effects of finite Hubbard interaction. 
    (a) and (b) AFM structure factor $S(\pi,\pi)$ and squared effective flux $\langle \Phi^2 \rangle$ as functions of $U$ at $g = 4.0$. The simultaneous discontinuous jumps in both quantities are the hallmark of a single first-order transition.
    (c) Correlation-length ratio $R_s$ as a function of $U$ at $g = 6.0$ for different system sizes. The crossing of curves for different $L$, shown in the inset, signals a continuous AFM transition at $U \approx 6.4t$.
    (d) Squared effective flux $\langle \Phi^2 \rangle$ at $g = 6.0$, showing a first-order photon condensation transition at $U\approx 8.6t$.
    (e) and (f) Histogram of probability distribution of the $S(\pi,\pi)$ at $U = 6.4$ (close to the AFM transition) and $\langle \Phi^2 \rangle$ at $U = 8.15$ (close to the superradiant transition) for $L=12$ system, the former displays a unimodal distribution characteristic of a continuous (second-order) phase transition, while the latter  exhibits a bimodal distribution characteristic of a first-order phase transition, $g$ is fixed as $g=6.0$ and the dashed line indicates the average value.
    (g) and (h) Finite-size scaling of the AFM structure factor $S(\pi,\pi)$ and the squared effective flux $\langle \Phi^2 \rangle$ for different system sizes at $g = 6.0, U=7.4$, confirming the coexistence of AFM order and photon condensation in the intermediate interaction regime.
    }
    \label{fig:fig3}
\end{figure}

In the weak coupling regime ($g < g_c = 2.8$) without photon condensation, the fermions exhibit a perfect fermi surface nesting at the momentum $(\pi, \pi)$, thus  an infinitesimal Hubbard interaction suffices to induce long-range antiferromagnetic order and drive the system into an insulating phase. Our primary focus is on the strong coupling regime ($g > g_c$). As depicted in the ground-state phase diagram in Fig.~\ref{fig:diagram}(b), an increasing Hubbard interaction \(U\) enhances the critical photon-electron coupling strength required for the photon condensation transition. Within an intermediate coupling regime $2.8<g<4.0$, we observe a single transition as a function of \(U\), where the AFM order develops concurrently with the destruction of photon condensation. As shown in Fig.~\ref{fig:fig3}(a) and (b), this transition is characterized by sharp discontinuities in both the AFM structure factor and the photon flux, indicating a first-order transition from a paramagnetic, superradiant chiral phase to an AFM phase without photon condensation.

More intriguingly, in the strong electron-photon coupling regime, onset of AFM ordering and the destruction of photon condensation no longer coincide, leaving an intermediate phase with coexistence of AFM order and photon condensation in the intermediate interaction regime, dubbed ``superradiant AFM" phase in Fig.~\ref{fig:diagram}(b). To illustrate this point, we fix \( g = 6.0 \),  and increase \( U \) from zero.  The system first undergoes a continuous AFM transition at \( U \approx 6.4t \). This is evidenced by the crossing of the correlation-length ratio $R_s$ for different system sizes [Fig.~\ref{fig:fig3}(c), inset] and a corresponding unimodal distribution in the histogram of $S(\pi,\pi)$ [Fig.~\ref{fig:fig3}(e)], confirming the second-order nature of the transition. As we further increase the interaction to \( U \approx 8.6t \), a  first-order photon condensation transition occurs [Fig.~\ref{fig:fig3}(d)]. Near this transition, the histogram of  $\langle \Phi^2\rangle$ displays a bimodal distribution [Fig.~\ref{fig:fig3}(f)], with peaks located at $\langle \Phi^2\rangle = 0$ and at a finite $\langle \Phi^2\rangle$, a hallmark of a first-order transition~\cite{SandvikNotes}. The intermediate regime bounded by these two critical points constitutes a phase where AFM order and photon condensation coexist, which can be further supported by the finite-size scaling analysis of their order parameters, as shown in Fig.~\ref{fig:fig3}(g) and (h) respectively.

\begin{figure}
    \centering
    \includegraphics[width=\linewidth]{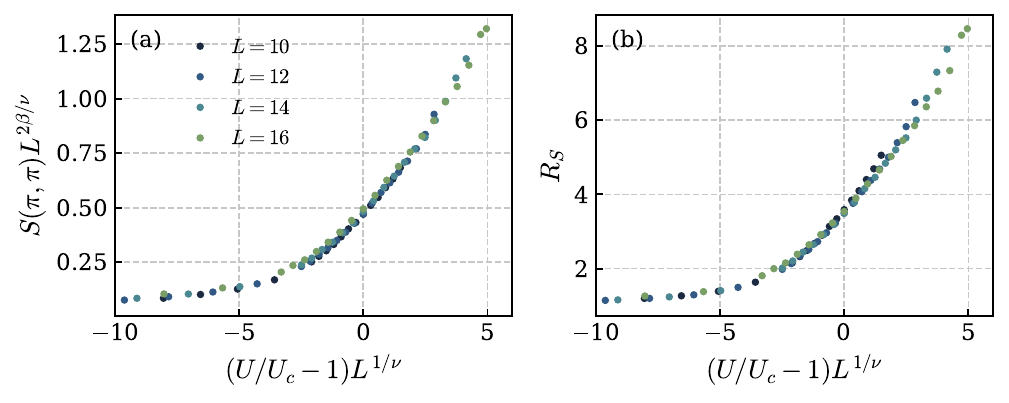}
    \caption{Finite-size scaling analysis of the continuous AFM transition at $g=6.0$. Data collapse of the (a) AFM structure factor $S(\pi,\pi)L^{2\beta/\nu}$ and (b) correlation-length ratio $R_s$ as functions of the scaling variable $(U/U_c-1)L^{1/\nu}$. The critical point here is $U_c\approx6.4$. The critical exponents used for the collapse, $\nu = 1.021, \beta=0.743$, are taken from the established values for the chiral Heisenberg Gross-Neveu universality class~\cite{Sorella16PRX}. }
    \label{fig:collapse}
\end{figure}

To elucidate the precise nature of this continuous transition from the superradiant chiral/$\pi$-flux phases to the superradiant AFM phase, we perform a finite-size scaling analysis, fixing $g$ and systematically varying the Hubbard interaction $U$. As illustrated in Fig.~\ref{fig:collapse}(a) and (b), our findings show that both the correlation-length ratio and the AFM structure factor exhibit a clear scaling collapse, aligning perfectly with the Gross-Neveu universality class~\cite{Rosenstein1993PLB,Herbut2006PRL,Assaad2013prx,li2017nc,Scherer2017PRD,Li2018SA,Lang2019prl,Yu2026SA,Li2024PRLNon-Hermitian,Sorella16PRX,Xu2026arXiv} with critical exponents adopted from Ref.~\cite{Sorella16PRX}. This transition is usually characterized by bosonic ordering accompanied by a gap opening in the Dirac spectrum. Our finite-size scaling results show that this universality class persists even in the presence of a dynamical instead of static magnetic flux, where the quantum fluctuation can make the flux per plaquette deviate from $\pi$.

\magenta{\it Experimental relevance|}   The proposed Hamiltonian in Eq.(\ref{model}) provides a minimum model for exploring magnetic photon condensation and the collective strong-coupling regime of interacting electrons coupled to a single cavity mode. In our simulation, we choose $\omega_c\sim t$ for numerical simplicity, while the photon frequency in a realistic cavity is much smaller than that ({\it e.g.} in copper oxides~\cite{Keimer2015}, the hopping amplitude $t$ is about $0.3$ eV$\sim$ 70THz), which can further facilitate the occurrence of superradiant phase transitions. The key challenge of the experimental realization of  superradiant physics within current optical cavities  is to access the deep strong-coupling regime ($g\sim \mathcal{O}(1)$): for a realistic cavity with $\omega_c=2\pi\times 300$GHz and resonance wavelength  $\lambda_c\sim 1\,\mu\mathrm{m}$, the value of $g$ can be estimated as $g\sim O(10^{-2})$, significantly smaller than the critical $g_c$ required to observe superradiant phase transition. Encouragingly, recent work has demonstrated ultra-strong magnetic coupling with a superconducting resonator~\cite{Ghirri2023}, underscoring the experimental viability of the physics captured in our model. Cold-atom systems offer another potential platform. However, because atoms are neutral, realizing our Hamiltonian would require the implementation of dynamical synthetic gauge fields~\cite{Kollath2016,Zheng2016}.

\magenta{\it Conclusion and Outlook|} In summary, we have established an unbiased numerical framework based on sign-problem-free QMC to explore the cavity strongly correlated electronic systems, and revealed the exotic quantum phases ({\it e.g.} the superradiant chiral/$\pi$-flux phase and superradiant AFM phases) and quantum phase transitions induced by the interplay between the cavity mode and repulsive interactions between fermions. Notably, the chiral superradiant phase identified here offers a highly controllable platform to explore the fractal spectrum of Hofstadter’s butterfly~\cite{Hofstadter1976PRB} and related strongly correlated phenomena, paralleling recent breakthroughs in  moir\'{e} architectures~\cite{Dean2013NatureHofstadter,Xu2025NP,Yazdani2025Nature,Feldman2022NP}. Our method has opened a new avenue for the non-perturbative study of strongly correlated cavity quantum materials, and is ready to be generalized to other systems such as the  cavity-embedded 2D materials and moir\'{e} superlattices~\cite{Kennes2021NPReview,Yao2023PRL,Yao2023PNAS,Finley2024SA,Du2023ScienceReview,Deng2021Nature},  cavity-mediated superconductivity or topological systems ({\it e.g.} a cavity Kane-Mele-Hubbard model~\cite{Hohenadler2011,Zheng2011}). 
    

{\it Acknowledgments} --- 
We thank Yu-Xiang Zhang and Tao Xiang for helpful discussions. K.W. appreciates Benjamin Cohen-Stead for assistance with the SmoQyDQMC package~\cite{SmoQyDQMC.jl}, Shaohang Shi for help with benchmarking, and Tao Xiang for encouragement and support of computational resources. Z.-X.L. is supported by the National Natural Science Foundation of China under Grant No. 12347107 and No. 12474146, Beijing Natural Science Foundation under Grant No. JR25007 and the New Cornerstone Investigator Program. Z.C. is supported by the National Key Research and Development Program of China (2024YFA1408303), Natural Science Foundation of China (Grant No.12525407),  Shanghai Municipal Science and Technology Major Project (Grant No.2019SHZDZX01), Shanghai Science and Technology Innovation Action Plan(Grant No. 24Z510205936).

\bibliography{ref}

\appendix
\begin{widetext}
\clearpage
\begin{center}
\textbf{\large Supplemental Material for \\ ``Superradiant strongly correlated quantum states in cavity Hubbard model"}
\end{center}

\section{The details of Determinant Monte Carlo and Hybrid Monte Carlo}
We begin by considering the Hamiltonian describing the light-matter interaction:
\begin{align}
\hat{\mathcal{H}} &= -t \sum_{\langle ij \rangle, \sigma} \left( e^{-i \hat{A}_{ij}} \hat{c}_{i\sigma}^\dagger \hat{c}_{j\sigma} + \text{h.c.} \right) 
+ U \sum_i \hat{n}_{i\uparrow} \hat{n}_{i\downarrow} 
+ \frac{\hat{P}^2}{2m} + \frac{1}{2} m \omega_c^2 \hat{X}^2.
\end{align}
Here, the coupling to the photon coordinate $\hat{X}$ is mediated by the Peierls phase operator $\hat{A}_{ij} = \int_{\mathbf{r}_i}^{\mathbf{r}_j} \hat{\mathbf{A}} \cdot d\mathbf{r}$. We adopt the symmetric gauge for the vector potential operator, defined as $\hat{\mathbf{A}} = \sqrt{2/N} g \hat{X}(-y/2, x/2)$ with $(x, y)$ being the spatial coordinates in the x-y plane.

At an inverse temperature $\beta = 1/T$, the partition function is expressed as:
\begin{equation}
Z = \operatorname{Tr}\left[ e^{-\beta \hat{\mathcal{H}}} \right].
\end{equation}
To evaluate this numerically, we discretize the imaginary time into $L_\tau$ slices of width $\Delta\tau = \beta / L_\tau$ and apply the Trotter–Suzuki decomposition to the propagator:
\begin{align}
\operatorname{Tr}\left[ e^{-\beta \hat{\mathcal{H}}} \right] = \operatorname{Tr}\left[\prod_l^{L_{\tau}} e^{-\Delta\tau \hat{\mathcal{H}}_{ph}} e^{-\Delta\tau \hat{\mathcal{H}}_U} e^{-\Delta\tau \hat{\mathcal{H}}_K[\hat{X}]} \right] + O(\Delta\tau^2). 
\end{align}
Here, $\hat{\mathcal{H}}_K[\hat{X}]$ represents the kinetic hopping term coupled to the cavity field via the Peierls phase, $\hat{\mathcal{H}}_U$ is the on-site Hubbard repulsion, and $\hat{\mathcal{H}}_{ph}$ describes the bare harmonic oscillator energy of the photon field. In our calculation, we employ a time step of $\Delta\tau = 0.1$, having verified that the resulting Trotter error is sufficiently small and does not affect the numerical results. 

We decouple the on-site interaction term using a discrete Hubbard–Stratonovich (HS) transformation in the density channel~\cite{Hirsch83,Hirsch85}:
\begin{align}
e^{-\Delta\tau U (\hat{n}_{i\uparrow}-\frac{1}{2})(\hat{n}_{i\downarrow}-\frac{1}{2})} 
= \frac{e^{{\Delta\tau U /4}}}{2} \sum_{s_i = \pm 1} e^{\lambda s_i (\hat{n}_{i\uparrow} + \hat{n}_{i\downarrow} - 1)}, 
\end{align}
with the coupling constant determined by $\cosh \lambda = \exp(-\Delta\tau U / 2)$. This transformation introduces binary Ising auxiliary fields $s_{i,l} \in \{\pm 1\}$ at each lattice site $i$ and imaginary time slice $l$. Although this decomposition involves a complex parameter $\lambda$—making the simulation computationally more demanding than real-variable schemes—it has the crucial advantage of explicitly preserving SU(2) spin-rotational symmetry for every configuration of the auxiliary field. This conservation law significantly suppresses statistical fluctuations, enabling a more precise determination of magnetic phase transitions.

Upon integrating out the fermionic degrees of freedom, the partition function takes the form:
\begin{align}
Z = \int \mathcal{D}[X_l] \sum_{\{s\}} e^{-S_{ph}[\{X\}]} \, w[\{s\}] \, \prod_{\sigma} \det \mathbf{M}_\sigma[\{X\}, \{s\}],
\end{align}
where the $\{X\}$ and $\{s\}$ are the configurations of time-dependent photon coordinate fields and space-time dependent Ising auxiliary fields, the weight resulting from the auxiliary field transformation is $w[s] = \frac{e^{{\Delta\tau U /4}}}{2} \exp\left(-\lambda \sum_{i,l} s_{i,l}\right)$, and the discretized action for the photon field in the photon coordinate representation is given by:
\begin{align}
S_{ph}[\{X\}] = \sum_{l=1}^{L_\tau} \Delta\tau \left[ \frac{m}{2} \left( \frac{X_{l+1} - X_l}{\Delta\tau} \right)^2 + \frac{1}{2} m \omega_c^2 X_l^2 \right].
\end{align}
The fermion matrix $\mathbf{M}_\sigma$ encodes the coupling to both the photon field $\{X\}$ and the Ising auxiliary field $\{s\}$. It is defined as
\begin{equation}
\mathbf{M}_\sigma = \mathbf{I} + \mathbf{B}_{L_{\tau}, \sigma} \mathbf{B}_{L_{\tau} - 1, \sigma} \cdots \mathbf{B}_{1, \sigma},
\end{equation}
where $\mathbf{I}$ is the identity matrix and the propagator for a single time slice $l$ is
\begin{equation}
\mathbf{B}_{l, \sigma} = e^{-\Delta\tau \mathbf{K}_\sigma[X_l]} e^{-\Delta\tau \mathbf{V}_\sigma[s_l]}.
\end{equation}
Here, $\mathbf{K}_\sigma[X_l]$ represents the kinetic energy matrix at time slice $l$, which incorporates the Peierls phase induced by the photon field. Its elements in real space are:
\begin{equation}
(\mathbf{K}_{\sigma}[X_l])_{ij} = 
\begin{cases}
-t\, e^{-i A_{ij}(X_l)} & \text{if } \langle i,j \rangle \text{ are nearest neighbors}, \\
0 & \text{otherwise}.
\end{cases}
\end{equation}
The potential matrix $\mathbf{V}_\sigma[\{s_l\}]$, resulting from the density-channel decoupling, is diagonal in real space:
\begin{equation}
(\mathbf{V}_{\sigma}[\{s_l\}])_{ij} = \lambda s_{i,l} \delta_{ij}, \quad \text{with} \quad \cosh(\lambda) = e^{-\Delta\tau U / 2}.
\end{equation}
This formulation allows for the efficient construction of the equal-time Green’s function via $\mathbf{G}_\sigma = \mathbf{M}_\sigma^{-1}$, which is essential for computing physical observables and determining the fermionic forces required for the Hybrid Monte Carlo updates.

\subsection*{Local update for Ising auxiliary fields}
The discrete HS field \( s_i(\tau) \) is updated via local Metropolis sampling. For each space-time point $(i,l)$ a flip \( s_{i,l} \to -s_{i,l} \) is proposed and accepted with probability:
\begin{align}
P_{\text{acc}} = \min \left[ 1, \frac{w[\{s'\}]\det \mathbf{M}_\uparrow[\{s'\}] \det \mathbf{M}_\downarrow[\{s'\}]}{w[\{s\}]\det \mathbf{M}_\uparrow[\{s\}] \det \mathbf{M}_\downarrow[\{s\}]} \right],
\end{align}
where $ \{s'\} $ is the flipped configuration. In our implementation, we employ the Sherman–Morrison formula to evaluate the determinant ratios and update the Green's function via fast rank-one updates. This reduces the computational cost of an accepted move from $O(N^3)$ to $O(N^2)$, where $N$ is the number of lattice sites. Consequently, the total computational complexity for a full sweep of the lattice scales as $O(N^3 L_\tau)$\cite{AssaadReview}.

\subsection*{Hybrid Monte Carlo for photon fields}
Standard local updates are inefficient for sampling the continuous photon field $\mathbf{X}$. To overcome this, we employ the Hybrid Monte Carlo (HMC) method~\cite{HQMC87,HQMCCM92,SmoQyDQMC.jl,Ben22PRE}, which utilizes global updates based on Hamiltonian dynamics. We introduce a fictitious conjugate momentum field $\Pi_l$ for each photon field $X_l$ and define a classical Hamiltonian $\mathcal{H}_{\text{cl}}$ governing the evolution in a fictitious time $t$:
\begin{align}
\mathcal{H}_{\text{cl}}[\{X\}, \{\Pi\}] = \sum_{l=1}^{L_\tau} \frac{1}{2} \Pi_l^2 + S_{ph}[\{X\}] - \sum_{\sigma} \ln \det \mathbf{M}_\sigma[\{X\}].
\end{align}
Here, the first term represents the classical kinetic energy, and the remaining terms constitute the effective potential energy derived from the bosonic action and the fermion determinants. The HMC procedure evolves the phase space variables $(X, \Pi)$ at each imaginary-time slice $l$ according to the equations of motion:
\begin{align}
\frac{dX}{dt} = \frac{\partial \mathcal{H}_{\text{cl}}}{\partial \Pi} = \Pi, \quad 
\frac{d\Pi}{dt} = -\frac{\partial \mathcal{H}_{\text{cl}}}{\partial X} = -F[X],
\end{align}
where $F[X]$ is the generalized force. We integrate these equations numerically using the reversible, symplectic leapfrog algorithm with a time step $\delta t$:
\begin{align}
\Pi(t + \delta t/2) &= \Pi(t) - \frac{\delta t}{2} \frac{\partial S_{\text{eff}}}{\partial X(t)}, \\
X(t + \delta t) &= X(t) + \delta t\, \Pi(t + \delta t/2), \\
\Pi(t + \delta t) &= \Pi(t + \delta t/2) - \frac{\delta t}{2} \frac{\partial S_{\text{eff}}}{\partial X(t + \delta t)}.
\end{align}
The resulting field configuration is accepted via a Metropolis step with probability $P = \min \left[1, \exp(-\Delta \mathcal{H}_{\text{cl}}) \right]$, where $\Delta \mathcal{H}_{\text{cl}}$ accounts for energy conservation errors introduced by the finite integration step size.

The total force acting on \( X \) consists of a bosonic and fermionic part:
\begin{align}
    \frac{\partial \mathcal{S}}{\partial X} = \frac{\partial \mathcal{S}_B}{\partial X} + \frac{\partial \mathcal{S}_f}{\partial X}.
\end{align}
The bosonic force is local:
\begin{align}
    \frac{\partial \mathcal{S}_B}{\partial X_l} = \Delta \tau\, m\left( \omega_c^2 X_l + \frac{2 X_l - X_{l+1} - X_{l-1}}{\Delta\tau^2} \right).
\end{align}

The fermionic part involves the derivative of the fermion determinant:
\begin{align}
    \frac{\partial \mathcal{S}_f}{\partial X} = -\sum_\sigma \text{Tr} \left[ \frac{\partial M_\sigma}{\partial X} M_\sigma^{-1} \right],
\end{align}
which, under Trotter decomposition \( M = B_{L_\tau} \cdots B_1 \), is explicitly written as:
\begin{align}
    \frac{\partial \mathcal{S}_f}{\partial X} 
    &= \text{Tr} \left[ \frac{\partial B}{\partial X} B^{-1}(G - I) \right] \nonumber\\
    &= \text{Tr} \left[ \frac{\partial e^{-\Delta \tau V}}{\partial X} e^{\Delta \tau V}(G - I) \right] 
    + \text{Tr} \left[ \frac{\partial e^{-\Delta \tau K}}{\partial X} e^{\Delta \tau K} e^{\Delta \tau V}(G - I) e^{-\Delta \tau V} \right] \nonumber\\
    &= -\Delta\tau\, \text{Tr} \left[ \frac{\partial V}{\partial X}(G - I) \right] 
    - \Delta\tau\, \text{Tr} \left[ \frac{\partial K}{\partial X} e^{\Delta\tau V}(G - I)e^{-\Delta\tau V} \right],
\end{align}
where \( G \equiv G_\sigma(\tau, \tau) \) is the equal-time Green’s function. Here, \( K \) and \( V \) are the kinetic and potential matrices in the split-operator representation. The dominant numerical cost lies in computing the trace terms involving the fermion matrix derivative and Green’s function.

\subsection*{Absence of Sign Problem at Half Filling}
The model is free from the sign problem at half-filling. To demonstrate this, we perform a partial particle-hole transformation acting exclusively on the spin-down fermions:
\begin{equation}
\hat{c}_{i\downarrow} \rightarrow (-1)^i \hat{d}_{i\downarrow}^\dagger, \quad \hat{c}_{i\downarrow}^\dagger \rightarrow (-1)^i \hat{d}_{i\downarrow},
\end{equation}
where the phase factor $(-1)^i$ is $+1$ for sites on sublattice A and $-1$ for sublattice B. Under this transformation, the particle number operator maps to $\hat{n}_{i\downarrow} \rightarrow 1 - \hat{\tilde{n}}_{i\downarrow}$, where $\hat{\tilde{n}}_{i\downarrow} = \hat{d}_{i\downarrow}^\dagger \hat{d}_{i\downarrow}$ is the number operator in the transformed basis.

Under this transformation, the kinetic energy term transforms as:
\begin{equation}
-t e^{-i A_{ij}} \hat{c}_{i\downarrow}^\dagger \hat{c}_{j\downarrow}  -t e^{i A_{ij}} \hat{c}_{j\downarrow}^\dagger \hat{c}_{i\downarrow} \rightarrow -t e^{i A_{ij}} \hat{d}_{i\downarrow}^\dagger \hat{d}_{j\downarrow} -t e^{-i A_{ij}} \hat{d}_{j\downarrow}^\dagger \hat{d}_{i\downarrow} .
\end{equation}
This implies that the kinetic energy matrix for the transformed spin-down sector, $\mathbf{K}_\downarrow$, is the complex conjugate of the spin-up matrix: $\mathbf{K}_\downarrow = \mathbf{K}_\uparrow^*$. Meanwhile, the exponent in the Hubbard–Stratonovich decoupling transforms as:
\begin{equation}
\lambda s_i (\hat{n}_{i\uparrow} + \hat{n}_{i\downarrow} - 1) \rightarrow \lambda s_i (\hat{\tilde{n}}_{i\uparrow} - \hat{\tilde{n}}_{i\downarrow}).
\end{equation}
Consequently, the effective potential matrix for the spin-down holes is $\mathbf{V}_\downarrow = -\mathbf{V}_\uparrow$. Since the repulsive Hubbard model requires a purely imaginary coupling $\lambda$ (such that $\lambda^* = -\lambda$), we obtain the relationship $\mathbf{V}_\downarrow = \mathbf{V}_\uparrow^*$. Combining these results, the single time-slice propagator for the spin-down sector satisfies:
\begin{equation}
\mathbf{B}_\downarrow(\tau) = e^{-\Delta\tau \mathbf{K}_\uparrow^*} e^{-\Delta\tau \mathbf{V}_\uparrow^*} = \left( e^{-\Delta\tau \mathbf{K}_\uparrow} e^{-\Delta\tau \mathbf{V}_\uparrow} \right)^* = \mathbf{B}_\uparrow(\tau)^*.
\end{equation}
This ensures that the product of determinants is real and non-negative ($\det \mathbf{M}_\uparrow \det \mathbf{M}_\downarrow = |\det \mathbf{M}_\uparrow|^2 \ge 0$), thereby eliminating the sign problem. Thus, we have proven the absence of the sign problem for the half-filled cavity-Hubbard model, provided the lattice is bipartite~\cite{lzx2019review,wu05sign,wei2016sign}.

\section{Additional Numerical Results and Discussion}

In this section, we present extended numerical data characterizing the phase diagram and the nature of the quantum phase transitions arising from the interplay between electronic correlations and cavity-mediated interactions. We track the evolution of the total ground-state energy $E_t$, the photon fluctuation order parameter $\langle \Phi^2 \rangle$, the antiferromagnetic (AFM) structure factor $S(\pi,\pi)$, and the correlation length ratio $R_s$ as a function of the Hubbard $U$ for distinct light-matter coupling strengths $g \in \{2.0, 4.0, 6.0, 10.0\}$.

\textbf{Weak Coupling Regime ($g=2.0$):} Since $g$ is below the critical threshold for photon condensation ($g < g_c \approx 2.8$), the system remains in the normal metallic phase with $\langle \Phi^2 \rangle \approx 0$ for $U=0$. The physics here is dominated by the electronic instability of the half-filled square lattice. As shown in the first row of Fig.~\ref{fig:S1}, the AFM structure factor $S(\pi,\pi)$ rises sharply upon introducing $U$ and the correlation-length ratio $R_s$ increases with system size in the entire region of interaction strength. This behavior aligns with the standard half-filled 2D Hubbard model, in which perfect nesting facilitates an instability toward magnetic ordering even at infinitesimal interactions.

\textbf{Intermediate Coupling Regime ($g=4.0$):} In this regime, the cavity triggers a photon-condensation phase characterized by a generated flux that breaks time-reversal symmetry, giving rise to a Hofstadter butterfly spectrum. This photon-condensed state competes strongly with the tendency toward Mott insulation. As $U$ increases, the energy penalty for double occupancy suppresses the charge fluctuations required to sustain the cavity-mediated kinetic energy gain. This competition drives a simultaneous first-order transition: the photon condensate collapses precisely when antiferromagnetic (AFM) order emerges. This is evidenced by a sharp discontinuity in the slope of the ground-state energy and abrupt jumps in both $\langle \Phi^2 \rangle$ and $S(\pi,\pi)$. 

\textbf{Strong Coupling Regime ($g=6.0$ and $10.0$):} In the strong-coupling regime, the cavity induces a chiral superradiant phase where the effective flux in each plaquette gradually approaches $\pi$ as the coupling increases. On a square lattice, the $\pi$-flux phase gives rise to emergent Dirac fermions at low energies. Consequently, the system is characterized by low-energy Dirac fermions interacting with quantum fluctuations of the flux. Notably, these emergent fermions exhibit enhanced stability against correlations. The onset of AFM order does not immediately destabilize the photon condensate; instead, we observe a broad coexistence region where AFM order develops within the superradiant phase. By breaking spin rotational symmetry, this AFM order opens a gap in the Dirac spectrum. As a result, the system undergoes two distinct phase transitions: a continuous Gross-Neveu transition associated with the magnetic order, followed by a first-order transition corresponding to the loss of the photon field.

Consistent with the nature of a continuous phase transition, the AFM correlation-length ratios for different system sizes intersect at a well-defined crossing point, allowing for a precise determination of the critical points. By comparing $g=10.0$ with $g=6.0$, we find that stronger light-matter coupling significantly stabilizes the photon condensate. While the onset of AFM order, $U_{c, \text{AFM}}$, remains comparable (as indicated by $R_s$ in the third and fourth rows of Fig.~\ref{fig:S1}), the critical interaction strength required to suppress the photon condensate, $U_{c, \text{ph}}$, shifts to a much higher value (as shown by $\langle \Phi^2 \rangle$ in the third and fourth rows of Fig.~\ref{fig:S1}). This suggests that in the ultra-strong coupling limit, the emergent photon condensation can coexist with AFM magnetic order over a broad range of interaction strengths.

\begin{figure}[hbt]
    \centering
    \includegraphics[width=1.0\linewidth]{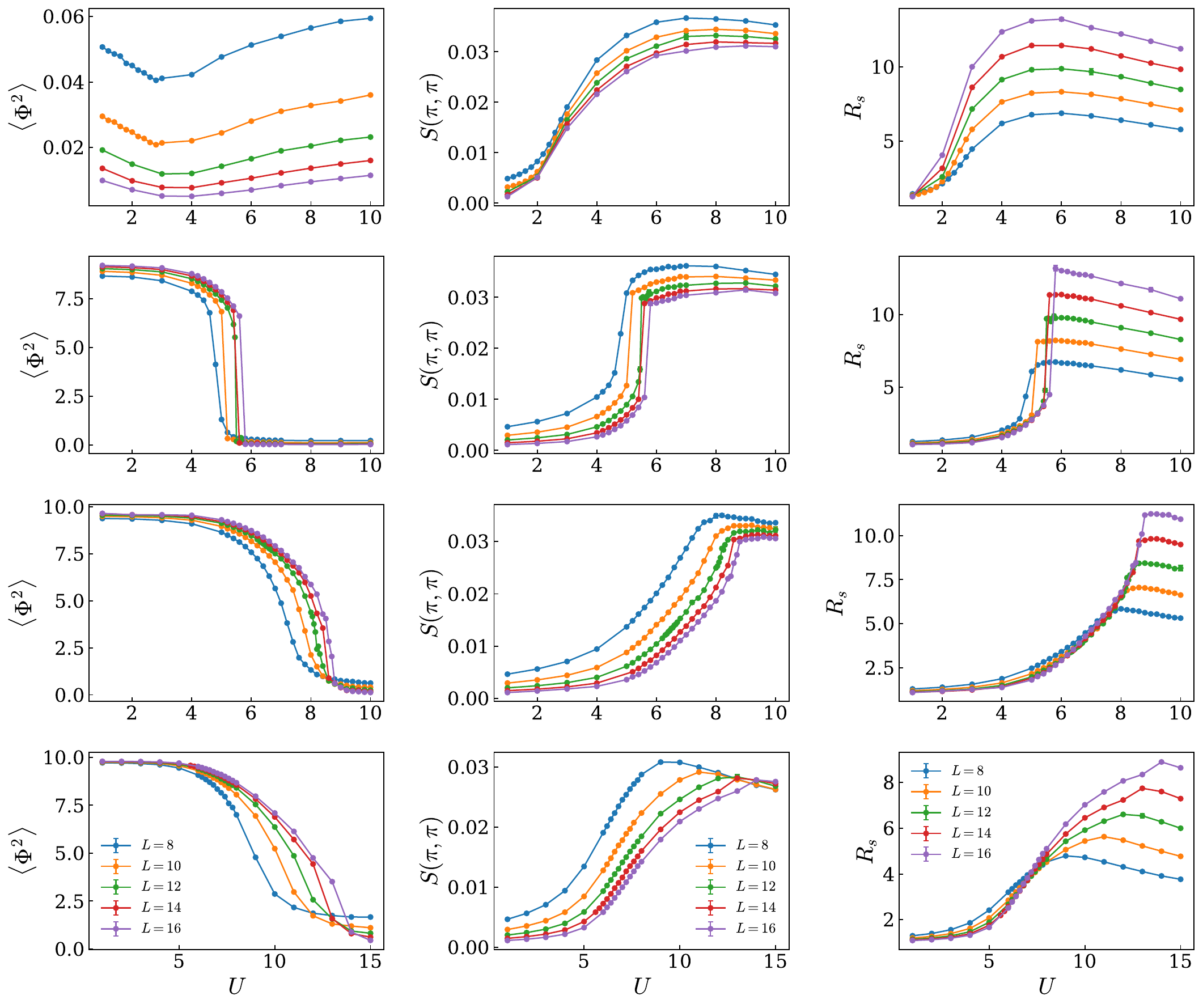}
    \caption{Evolution of physical observables as a function of interaction strength $U$ for different light-matter coupling values $g=2.0, 4.0, 6.0,$ and $10.0$ (arranged from top to bottom). From left to right, each column displays: the squared effective flux $\langle\Phi^2\rangle$, the antiferromagnetic spin structure factor $S(\pi,\pi)$, and the dimensionless correlation-length ratio $R_s$. These results provide further evidence for the competition between magnetic and photonic orders across different parameter regimes. }
    \label{fig:S1}
\end{figure}

\section{Detailed Analysis of Phase Transitions}

In this section, we present a detailed numerical analysis of the nature of the phase transitions in the strong coupling regime ($g=6.0$), where the antiferromagnetic (AFM) and photon condensation transitions decouple. As discussed in the main text, while the intermediate coupling regime ($g=4.0$) exhibits a clear-cut simultaneous first-order transition with unambiguous discontinuities in both order parameters, the strong coupling regime requires more careful analysis since the two transitions occur at different critical points. To rigorously determine the order of each transition, we analyze the probability distribution histograms of the relevant order parameters near the critical points. A bimodal distribution (double-peak structure) in the histogram is a hallmark of a first-order phase transition, indicating the coexistence of two distinct phases at the critical point. Conversely, a unimodal distribution that shifts continuously signals a second-order (continuous) transition~\cite{SandvikNotes}.

At strong coupling $g=6.0$, the transitions decouple: the AFM order sets in at a lower interaction strength ($U_{c_1} \approx 6.4$), while the photon condensation persists until a higher interaction strength ($U_{c_2} \approx 8.2$ for the $L=12$ system used in the histogram analysis).

\textit{Continuous AFM Transition:} Fig.~\ref{fig:S3_Sz} shows the histograms of the AFM structure factor $S(\pi,\pi)$ across the magnetic transition. We observe a unimodal distribution for all values of $U$. The single peak shifts continuously from zero to finite values as $U$ increases. The absence of a double-peak structure confirms that the AFM transition at $g=6.0$ is a continuous quantum phase transition. This result is consistent with the finite-size scaling analysis in the main text, which identifies this transition as a continuous one belonging to the Gross-Neveu universality class.

\begin{figure}[h]
\centering
\includegraphics[width=0.95\linewidth]{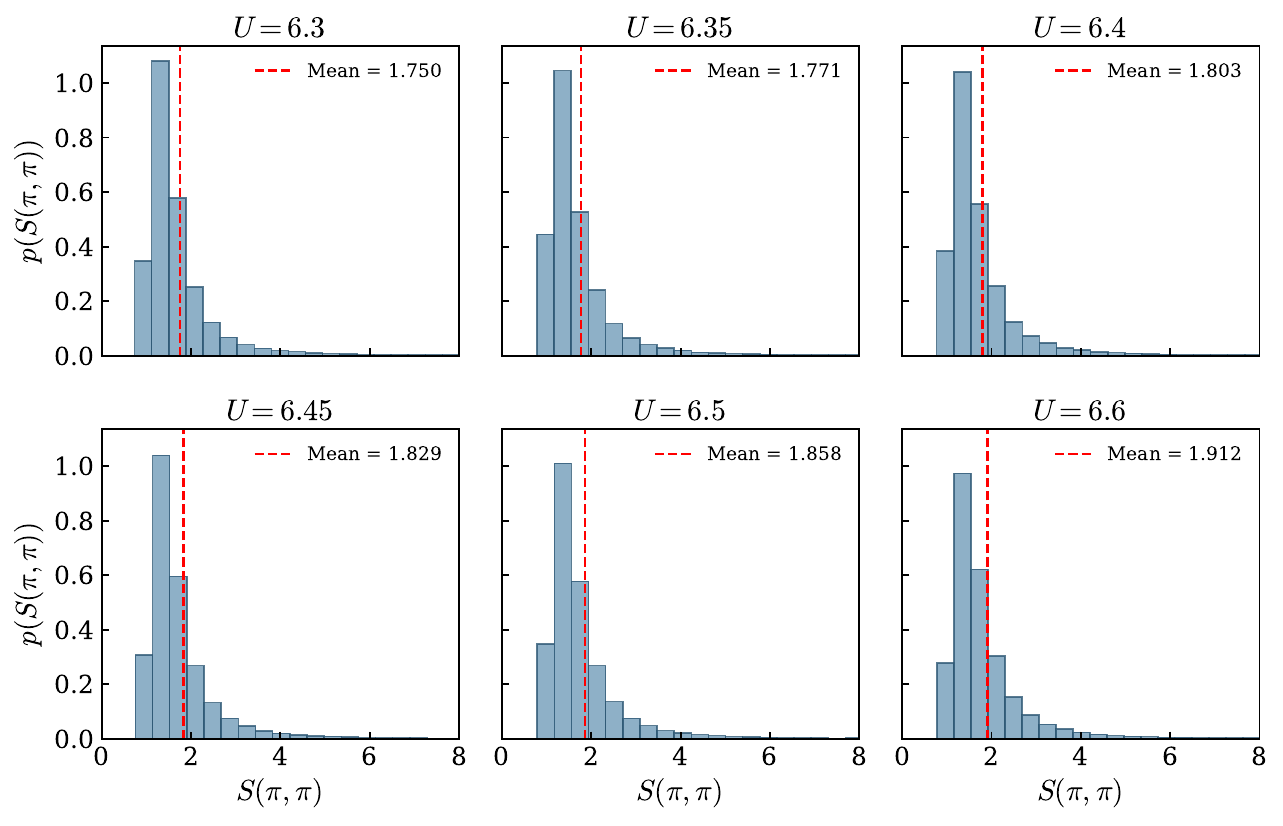}
\caption{Probability distribution of the AFM structure factor $S(\pi,\pi)$ at $g=6.0$. The system size is $L=12$. The distributions remain unimodal across the critical region ($U \approx 6.4$), shifting smoothly to higher values. This indicates a continuous phase transition.}
\label{fig:S3_Sz}
\end{figure}

\textit{First-Order Photon Condensation:} Fig.~\ref{fig:S4_X2} displays the histograms of the squared photon flux $\langle \Phi^2 \rangle$  near the second critical point $U_{c_2} \approx 8.2$ for the $L=12$ system. Here, a distinct bimodal structure emerges in the critical region (e.g., at $U=8.2$ and $U=8.25$), where peaks corresponding to the condensed phase (large $\langle \Phi^2 \rangle$) and the normal phase (small $\langle \Phi^2 \rangle$) coexist. This confirms that the destruction of photon condensation remains a first-order transition, even when decoupled from the magnetic onset.

\begin{figure}[h]
\centering
\includegraphics[width=0.95\linewidth]{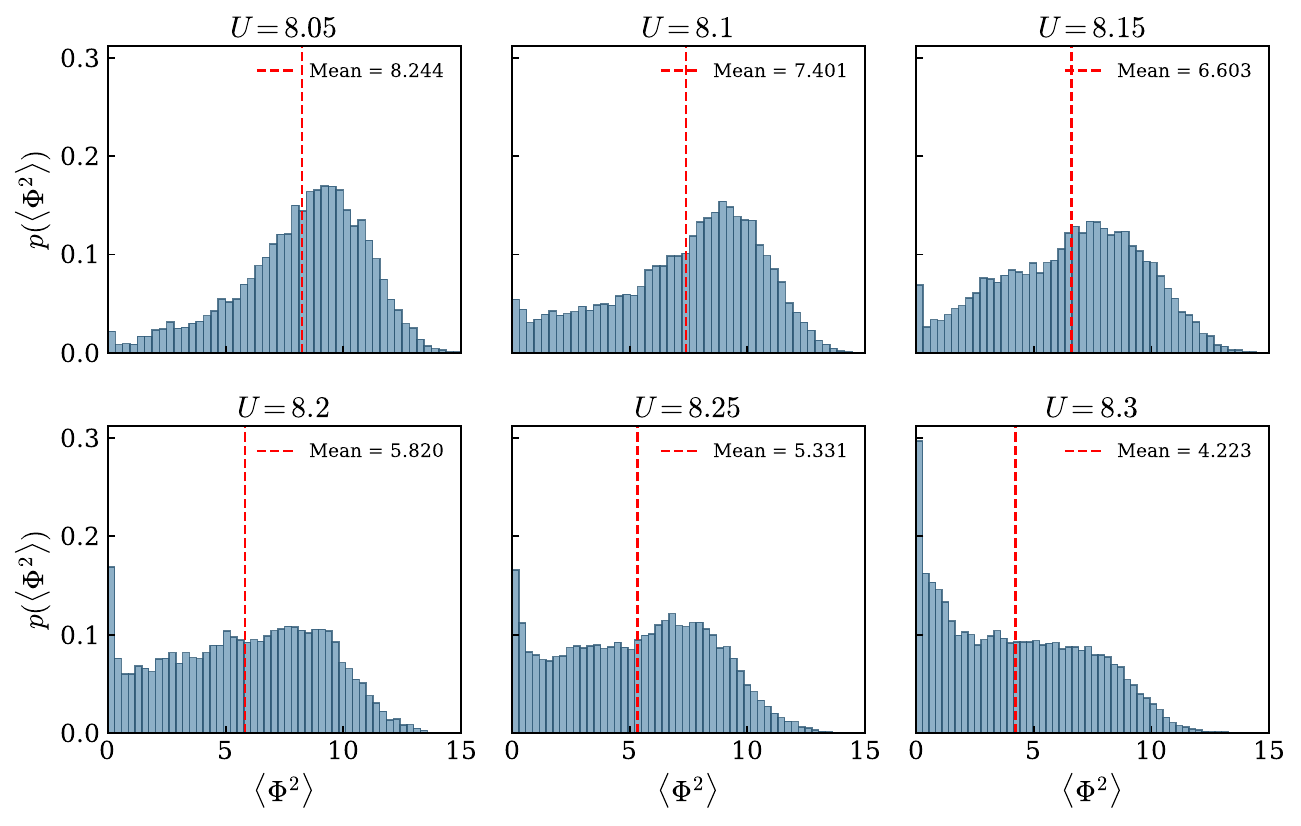}
\caption{Probability distribution histogram of $\langle \Phi^2 \rangle$ at $g=6.0$. Near the critical point $U \approx 8.2$ for the $L=12$ system, a clear bimodal distribution is observed (coexistence of high and low values), indicating the first-order nature of the photon condensation transition.}
\label{fig:S4_X2}
\end{figure}

\section{Initial Condition and Thermalization Analysis}
\label{sec:initial_condition}

Near a first-order phase transition, the free energy landscape exhibits multiple local minima separated by energy barriers. For each simulation, we ensure sufficient thermalization such that the measured observables reach a stationary value. However, convergence to a stationary state does not guarantee convergence to the true ground state---the system may remain trapped in a metastable local minimum. To resolve this ambiguity, we perform parallel simulations initialized from distinct photon field configurations and identify the thermodynamic ground state as the one with the lowest converged energy, as shown in Fig.~\ref{fig:S5Initial}.

As illustrated in Fig.~\ref{fig:S5Initial}, far from the transition, the observable $\langle \Phi^2 \rangle$ converges to the same value regardless of the initial conditions. However, in the vicinity of the transition point $g_c$, the system may instead converge to distinct metastable states. The resulting hysteresis in $\langle \Phi^2 \rangle$ is a direct consequence of the first-order nature of the phase transition.

\begin{figure}[h]
    \centering \includegraphics[width=0.9\linewidth]{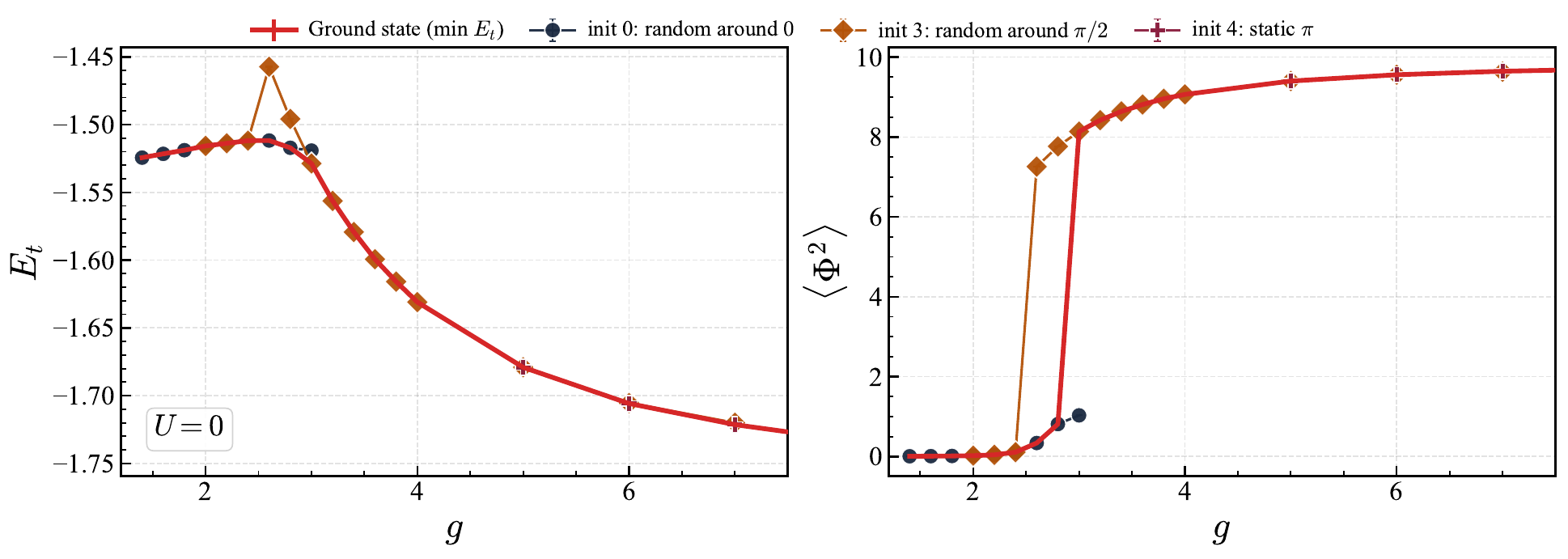}
    \caption{Initial-state dependence and ground-state selection. 
Comparison of QMC results for the total energy $E_t$ (left column) and squared photon flux $\langle \Phi^2 \rangle$ (right column) for system size $L=12$. 
Panels (a) and (b) show the evolution with light-matter coupling $g$ at fixed $U=0$. 
Different symbols correspond to QMC simulations initialized from distinct flux configurations, revealing metastable branches in the vicinity of the transition. 
The solid red lines trace the state with the lowest total energy and thus identify the thermodynamic ground state. 
Their crossings with competing branches signal the first-order transition points.}
\label{fig:S5Initial}
\end{figure}

\end{widetext}

\end{document}